\documentclass[12pt]{article}

\newcommand{\eq}{\begin{equation}}
\newcommand{\eqa}{\begin{eqnarray}}
\newcommand{\en}{\end{equation}}
\newcommand{\ena}{\end{eqnarray}}
\newcommand{\enn}{\nonumber \end{equation}}

\def\sk{\vskip .4cm}
\def\noi{\noindent}
\def\om{\omega}
\def\al{\alpha}
\def\be{\beta}

\def\Ga{\Gamma}

\let \si\sigma
\let \part\partial

\def\unmezzo{{1 \over 2}}
\def\epsi{\varepsilon}

\def\de{\delta}

\def\tv{{\bf t}}

\def\part{\partial}

\def\sk{\vskip .4cm}

\def\noi{\noindent}

\def\X0{X^0}

\def\om{\omega}

\def\al{\alpha}

\def\unmezzo{{1 \over 2}}
\def\epsi{\varepsilon}

\def\psib{{\bar \psi}}

\def\de{\delta}
\def\CABC{{C^A}_{BC}}

\def\tbo{{\bf t}}

\def\c#1#2{ C_{~#1}^{#2} }
\def\cl#1#2{ C_{~#1}^{#2} }

\def\square{{\,\lower0.9pt\vbox{\hrule \hbox{\vrule height 0.2 cm
\hskip 0.2 cm \vrule height 0.2 cm}\hrule}\,}}

\def\xtilde{{\tilde x}}
\def\ptilde{{\tilde p}}
\def\sitilde{{\tilde \sigma}}

\def\atilde{{\tilde a}}
\def\btilde{{\tilde b}}

\begin{document}

\begin{titlepage}
 \vskip 2em
\begin{center}
{\Large \bf  Higher form gauge fields and their nonassociative symmetry algebras }
\\[3em]
{\large {\bf Leonardo Castellani} } \\ [2em] {\sl Dipartimento di Scienze e Innovazione Tecnologica
\\Universit\`a del Piemonte Orientale,\\and  INFN Sezione di Torino\\ Viale T. Michel 11,  15121 Alessandria, Italy
}\\ [4em]
\end{center}

\begin{abstract}

We show that geometric theories with $p$-form gauge fields have a nonassociative symmetry structure, extending an underlying Lie algebra. This nonassociativity is controlled by the same Chevalley-Eilenberg cohomology that classifies free differential algebras, $p$-form
generalizations of Cartan-Maurer equations. A possible relation with flux backgrounds of closed string theory is pointed out.

\end{abstract}

\vskip 9cm \noi \hrule \vskip.2cm \noi {\small
leonardo.castellani@mfn.unipmn.it}

\end{titlepage}

\newpage
\setcounter{page}{1}

Free differential algebras (FDA's) \cite{fda1,fda0,fda2} (for reviews see for ex.\cite{gm2,gm3}) provide a convenient algebraic setting
for field theories with antisymmetric tensors. They generalize the Cartan-Maurer equations satisfied by the 1-form vielbeins on a group manifold $G$ by including $p$-form fields, and have been extensively used in the construction of (super)gravity theories. Given a FDA, there is a well-defined procedure to obtain a lagrangian for a field theory containing the $p$-forms of the FDA as fundamental fields. 

 Their dual formulation, based on the algebra of Lie derivatives on the ``FDA manifold",  has been studied in 
ref.s \cite{fda3,fda4,fda5}, and uses an extended Lie derivative along these antisymmetric tensors.  

In the present paper we show that the resulting algebra of Lie derivatives 
is nonassociative, and that the nonassociativity is controlled by the same
Chevalley-Eilenberg cohomology that classifies FDA's. In fact, our interest in studying
the nonassociativity of FDA dual algebras has been prompted by a recent paper \cite{BL}, where flux backgrounds in closed string theory are described by nonassociative structures in double phase space, controlled again by Chevalley-Eilenberg cohomology (for a very partial list of references see for example
 \cite{Lust2010,Mylonas2012,Hohm2013,Blumenhagen2013}). Since flux backgrounds involve $p$-forms, it seems that
algebraic structures describing $p$-forms tend to exhibit nonassociativity, depending
on nontrivial cohomology classes,  both for flux backgrounds
and in FDA's dual algebras. A clue for a
possible relation is given in the example i) (see below), where the nonassociative
algebra of the R-flux model in \cite{Lust2010,BL} is identified with a particular FDA 
dual algebra.

\sk
\noi {\bf Free differential algebras}
\sk
 Consider the case of ordinary Cartan-Maurer one-forms $\sigma^A$ supplemented by a single $p$-form
$B^i$  in a representation $D^i_{~j}$ of $G$:
 \eqa & & d
\sigma^A + \unmezzo \CABC ~\sigma^B \sigma^C = 0 \label{fdaLie}\\
& & dB^i+C^i_{~Aj} \sigma^A B^j + {1 \over (p+1)!}
C^i_{~A_1...A_{p+1}} \sigma^{A_1}... \sigma^{A_{p+1}}\nonumber \\
 & & ~~~~~~~~~~~~~\equiv \nabla B^i +  {1 \over
(p+1)!} C^i_{~A_1...A_{p+1}} \sigma^{A_1}... \sigma^{A_{p+1}} =0
\label{fdaB} \ena
\noi  Products between forms are understood to be exterior products. Taking the exterior derivative of the left-hand sides 
of (\ref{fdaLie}), (\ref{fdaB}) and requiring $d^2=0$ translates into a set of conditions for the structure constants:
\eqa & & C^A_{~~B[C} C^B_{~~DE]} =0
\label{jacobi1} \\
   & &
C^i_{~Aj} C^j_{~Bk} - C^i_{~Bj} C^j_{~Ak} = C^C_{~AB} C^i_{~Ck}
\label{jacobi2} \\ & & 2 ~C^i_{~[A_1j} C^j_{~A_2...A_{p+2}]}
-(p+1) C^i_{~B[A_1...A_p} C^B_{~A_{p+1}A_{p+2}]}=0
 \label{jacobi3} \ena

\noi Eq.s (\ref{jacobi1}) are the usual Jacobi identities for the
Lie algebra $G$. Eq. (\ref{jacobi2}) implies that $(C_A)^i_{~j}
\equiv C^i_{~Aj}$ is a matrix representation of $G$, while eq.
(\ref{jacobi3}) states that $C^i \equiv C^i_{~A_1...A_{p+1}}
\sigma^{A_1}... \sigma^{A_{p+1}}$ is a $(p+1)$-cocycle, i.e.
$\nabla C^i = 0$. Clearly $C^i$ differing by covariantly exact pieces $\Phi^i$ ($\nabla \Phi^i = 0$) lead to equivalent FDA's via the redefinition $B^i \rightarrow B^i + \Phi^i$.
Thus nontrivial FDA extensions of the Lie algebra are classified by
nontrivial cohomology classes of the covariant derivative $\nabla$,
i.e. by Chevalley cohomology.

As an example we recall the FDA of D=11 supergravity \cite{fda1}:
\eqa
 & & d\om^{ab}-\om^{ac} \om^{cb}=0~~~[=R^{ab}]\nonumber\\
 & & dV^a-\om^{ab}V^b-{i\over 2}\psib \Gamma^a \psi =0~~~[=R^a]\nonumber\\
 & & d\psi - {1\over 4} \om^{ab} \Gamma^{ab} \psi = 0 ~~~[=\rho]\nonumber\\
 & & dA - {1\over 2} \psib \Gamma^{ab} \psi V^a V^b
 =0~~~[=R(A)]\label{FDAd11}
 \ena

 \noi (all index contractions with the $D=11$ Minkowski metric). The D=11 Fierz identity $\psib \Gamma^{ab}\psi\psib \Gamma^a \psi =0$
 ensures the FDA closure ($d^2=0$). Its Lie algebra part is
 the D=11 superPoincar\'e algebra, whose fundamental fields (corresponding to the Lie algebra generators
 $P_a, J_{ab}, Q$)
 are the vielbein $V^a$, the spin connection $\om^{ab}$ and the gravitino $\psi$. The
 3-form $A$ is in the identity representation of the Lie algebra,
 and thus no $i$-indices are needed.
 The structure constants $\cl{A_1...A_{p+1}}{i}$ of (\ref{fdaB})
 are in the present case given by $\cl{\al\be a b }{}=-12
 (C\Gamma_{ab})_{\al\be}$ (no upper index $i$),
 while the $\c{Aj}{i}$ vanish.
This FDA extension exists thanks to the nontrivial
4-cocycle $ \psib \Gamma^{ab} \psi V^a V^b$. The curvatures $R^{ab}$, $R^a$, $\rho$ and $R(A)$ (i.e. the right-hand sides of the FDA equations, nonvanishing for general field configurations outside the vacuum\footnote{in this case we speak of a ``soft FDA manifold."})
 are respectively the Lorentz curvature, the torsion, the gravitino field strength and the 3-form field strength. 
The lagrangian of $d=11$ supergravity can be written as a 11-form, made out of exterior products of fields and curvatures, and is therefore by construction  invariant under diffeomorphisms. Infinitesimal diffeomorphisms are generated by Lie derivatives along all the soft group manifold directions, and comprehend all the invariances of the theory. This is essentially the core of the group geometric method developed many years ago \cite{gm1,gm2,gm3}.

\sk
\noi {\bf Extended Lie derivatives}
\sk

Consider the Lie derivatives $\ell_{\epsi^A \tbo_A}$ along a generic tangent vector $\epsi^A \tbo_A$, where the $\tbo_A$ are a basis of tangent vectors on $G$ dual to the left invariant one-forms $\si^A$. By means of the FDA equations and the Cartan formula
  \eq
 \ell_{\epsi^A \tbo_A}    =   i_{\epsi^A \tbo_A} d + d~
i_{\epsi^A \tbo_A} \label{Lieder}
 \en
\noi one finds their action on the FDA forms:
\eqa
 & & \ell_{\epsi^B \tbo_B}  \si^A = d(i_{\epsi^B \tbo_B}\si^A)
       +i_{\epsi^B \tbo_B} d \si^A  \nonumber \\
        & & ~~~~~~~~~~ =  d \epsi^A - \CABC \epsi^B \si^C        \label{diffgroup}\\
& & \ell_{\epsi^B \tbo_B} B^i = d(i_{\epsi^B \tbo_B}B^i)
       +i_{\epsi^B \tbo_B} d B^i  \nonumber \\
& & ~~~~~~~~~~ = - C^i_{Bj}  \epsi^B B^j 
-\frac{1}{p!}C^i_{BA_1...A_p}  \epsi^B\si^{A_1}...\si^{A_p}
 \label{diffB1}
\ena
\noi {\it Note:}  The action of  the contraction $i_{\tbo_B} $ on the ``cotangent  FDA basis"  $\si^A, B^i$ is given by 
\eq 
i_{\tbo_B} \si^A=\de^A_B,~~~i_{\tbo_B}B^i=0
\en
\sk

Computing the commutator of two Lie derivatives  
on $\si^A$ and $B^i$ yields \cite{fda3,fda4,fda5}:
  \eqa & &
 \left[ \ell_{ \epsi^A_1 \tbo_A},\ell_{
\epsi^B_2 \tbo_B} \right] = \ell_{ \left[ \epsi^A_1 \partial_A
\epsi^C_2 - \epsi^A_2 \partial_A \epsi^C_1 +\epsi^A_1 \epsi^B_2
C^C_{AB} \right] \tbo_C} \nonumber \\
 & &~~~~~~~~~~~~~~~~~
 +{1 \over (p-1)!} ~\ell_{  \epsi^A_1 \epsi^B_2~ C^i_{ABA_1...A_{p-1}}
 \si^{A_1}...\si^{A_{p-1}}\tbo_i }
\label{FDAalgebra1} 
 \ena
\noi where the second line involves an {\it extended Lie derivative} defined as follows \cite{fda3}. First
introduce a new contraction operator $i_{\epsi^j \tv_j}$,
acting on a generic form $\om = \om_{i_1...i_n
A_1...A_m} B^{i_1}  ... B^{i_n}  \si^{A_1}  ...
\si^{A_m}$ as
\eq i_{\epsi^j \tv_j} \om = n~ \epsi^j \om_{j
i_2...i_n A_1...A_m} B^{i_2} ... B^{i_n}  \si^{A_1}  ...
\si^{A_m} \label{newcontraction}
\en
\noi {\sl where} $\epsi^j$ {\sl is a (p -1)-form}. This operator
still maps $p$-forms into $(p -1)$-forms.  We can also define the
contraction $i_{\tv_j}$, mapping $n$-forms into $(n-p)$-forms, by
setting \eq i_{\epsi^j \tv_j}=\epsi^j i_{\tv_j}
\en
\noi In particular
 \eq
 i_{\tv_j} B^i=\de^i_j,~~~i_{\tbo_j} \si^A =0
\en
\noi so that $\tv_j$ can be seen as the ``tangent vector" dual to
$B^j$. 

Note that $i_{\epsi^j \tv_j}$ vanishes on forms that do not
contain at least one factor $B^i$.  Then the extended Lie derivative
is defined by:
 \eq
 \ell_{\epsi^i \tv_i} \equiv i_{\epsi^i \tv_i}d + d~
i_{\epsi^i \tv_i} \label{newLie}
 \en
It commutes with $d$, satisfies the
Leibniz rule, and can be verified to act on the fundamental
fields as
 \eqa &
&\ell_{\epsi^j \tv_j}\si^A = 0 \label{newLieonmu}\\
& &\ell_{\epsi^j \tv_j}B^i = d\epsi^i + C^i_{~Aj} 
\si^A  \epsi^j \label{newLieonB} \ena
\noi by using the FDA equations 
 (\ref{fdaLie}), (\ref{fdaB}). For example if the parameter $(p -1)$-form is given by
$ \si^{A_1} ...\si^{A_{p-1}}$:
 \eqa
  & & \ell_{\si^{A_1} ...\si^{A_{p-1}}\tv_j}B^i = d(\si^{A_1} ...\si^{A_{p-1}}) +  C^i_{~Aj} \si^A \si^{A_1} ...\si^{A_{p-1}} = \nonumber \\
 & & = - ({p-1 \over 2} \c{B_0 B_1}{[A_1} \de^{A_2 ...A_{p-1}]}_{B_2 ...B_{p-1}} \de^i_j - 
\c{B_0j}{i} \de^{A_1 ...A_{p-1}}_{B_1 ...B_{p-1}}) \si^{B_0} ...\si^{B_{p-1}}
 \ena
Using (\ref{newLieonmu}) and (\ref{newLieonB}) and the Jacobi identities (\ref{jacobi1})-(\ref{jacobi3}) it is straightforward to 
compute the commutators between ordinary and extended Lie derivatives,
and between extended Lie derivatives:
\eqa
& &\left[ \ell_{ \epsi^A
\tbo_A} , \ell_{ \epsi^j \tbo_j} \right] = \ell_{[ \ell_{ \epsi^A
\tbo_A} \epsi^k + C^k_{Bj} \epsi^B
\epsi^j ] \tbo_k}\label{FDAalgebra2}\\ & &\left[ \ell_{ \epsi^i_1
\tbo_i} , \ell_{ \epsi^j_2 \tbo_j} \right] = 0
\label{FDAalgebra3}\ena
According to eq. (\ref{FDAalgebra1}) the commutator of two ordinary Lie derivatives {\sl contains
 an extra piece proportional to an extended Lie derivative}, that acts nontrivially only on the $p$-form $B^i$. Then, if computed on $\si^A$, the commutator in 
(\ref{FDAalgebra1}) closes on the usual Lie algebra. Only its action on $B^i$ reveals the extra term involving the extended Lie derivative.
\sk
\noi {\bf The FDA dual algebra}
\sk

By taking constant $\epsi$
parameters (and nonvanishing only for given directions) the commutations  (\ref{FDAalgebra1}), (\ref{FDAalgebra2})  and (\ref{FDAalgebra3}) become :
 \eqa & &[ \ell_{\tv_A},
\ell_{\tv_B}] = \c{AB}{C} \ell_{\tv_C} + {1 \over( p-1)!}
~\cl{ABA_1...A_{p-1}}{i} ~ \ell_{\sigma^{A_1}...\sigma^{A_{p-1}}
\tv_i}\label{dual1} \\
 & &[\ell_{\tv_A},
\ell_{\si^{B_1}...\si^{B_{p-1}} \tbo_i}] = [\c{Ai}{k}
\de^{B_1...B_{p-1}}_{C_1...C_{p-1}} - (p-1) \c{AC_{1}}{[B_1}
\de^{B_2...B_{p-1}]}_{C_2...C_{p-1}}\de^k_i]
\ell_{\si^{C_1}...\si^{C_{p-1}} \tbo_k}\label{dual2} \\ &
&[\ell_{\si^{A_1}...\si^{A_{p-1}}
\tbo_i},\ell_{\si^{B_1}...\si^{B_{p-1}}\tbo_j}]=0 \label{dual3}
 \ena
 \noi This algebra can be considered the dual of the FDA system given in
(\ref{fdaLie}), (\ref{fdaB}), and extends the $G$ Lie algebra of
ordinary Lie derivatives along the $\tbo_A$ tangent vectors of the $G$ group manifold.
Notice the essential presence of the
$(p-1)$-form $\si^{A_1}...\si^{A_{p-1}}$ in front of the ``tangent
vectors" $\tbo_i$. 

As a concrete example, in the case of $D=11$ supergravity the algebra  (\ref{dual1})-(\ref{dual3}) is given in ref. \cite{fda4}, and contains the supertranslation algebra with central charges discussed for ex. in ref.s \cite{BS,Town}.

In the following we use the simplified notations
 \eq
 T_A \equiv \ell_{\tv_A},~~~T^{A_1...A_{p-1}}_{i} \equiv \ell_{\si^{A_1}...\si^{A_{p-1}}
\tbo_i}
 \en
\sk
\noi {\bf Nonassociativity}
\sk

The dual algebra of a FDA is in general
{\it nonassociative}. Indeed computing the Jacobiator of three (usual) Lie derivatives,
and using the expressions (\ref{dual1}), (\ref{dual2}) for the commutators, yields:
 \eqa
 & &
 J_{ABC} \equiv[ [T_A,T_B], T_C] + cyclic~in~ABC= \nonumber\\
& & = {1 \over (p-1)!}((p-1) \cl{EA_2...A_{p-1}[AB}{i}\c{C]A_1}{E}- \cl{A_1A_2...A_{p-1}[AB}{j}\c{C] j}{i})T^{A_1 ...A_{p-1}}_i \label{Jac}
 \ena
\noi which is nonvanishing in general\footnote{ The nonassociativity of the dual algebra  (\ref{dual1})-(\ref{dual3}) 
was not recognized in ref.s \cite{fda3,fda4,fda5}.} (it is {\it not} proportional to the left-hand side of the Jacobi identities (\ref{jacobi3})). This is the main result of the present paper. Since Lie derivatives generate infinitesimal diffeomorphisms, and these are the symmetries of the theory
based on the FDA, it follows that {\it the FDA symmetries close a nonassociative algebra}.

It is worthwhile to check this result in the FDA $D=11$ supergravity example,
containing a 3-form $A$ in the identity representation of $G$ = superPoincar\'e group in 11 dimensions. Computing for example the Jacobiator (\ref{Jac}) for two supersymmetry transformations 
generated by the charges $Q_\alpha$ and $Q_\beta$ and a Lorentz rotation generated by
$M_{ab}$ yields
 \eq
  J_{\al \be [ab]} = \eta_{c[a} (C \Ga_{b]d} )_{\al\be} T^{cd}
 \en
with $T^{cd} \equiv \ell_{V^c V^d \tbo }$. This Jacobiator does not vanish when applied to the 3-form $A$, since $ T^{cd} A=
\ell_{V^c V^d \tbo } A = d(V^c V^d)$.

In the following we consider three other examples of nonassociative algebras originating from the dual formulation of FDA's. 

\sk
\sk
\noi {\bf Examples}
\sk
 i) $p=2$ and $B$ in a 1-dimensional representation (with a single-valued index $i=\bullet$). Then the dual FDA algebra is
 \eqa
 & & [T_A,T_B] = \c{AB}{C} T_C + C^\bullet_{ABC} T^C_\bullet \label{p2one} \\
& & [T_A, T^B_\bullet] = - \c{AC}{B} T^C_\bullet  + \c{A\bullet}{\bullet} T^B_\bullet\label{p2two} \\
  & &[T^A_\bullet, T^B_\bullet] =0 \label{p2three}
 \ena
with a corresponding Jacobiator
  \eq
  J_{ABC} = 3 (C^\bullet_{D[AB} \c{C]E}{D} - \c{E[AB}{\bullet}\c{C]\bullet}{\bullet}) T^E_\bullet \label{p2Jac}
 \en
Naming the generators as
  \eq
  T_A = (x_a, \xtilde_a, q),~~~~T^B_\bullet = (p^b,\ptilde^a, q')
  \en
 and taking as values of the structure constants 
 \eq
 \c{qa}{b} = \de^b_a, ~~~~\c{q\atilde}{\btilde} = \de^\btilde_\atilde,~~~~C^\bullet_{abc}=R_{abc},~~~~\c{q\bullet}{\bullet}=3 \label{sc}
 \en
 \noi (all other structure constants vanishing, the indices $\atilde$ labeling the $\xtilde$ or $\ptilde$ directions, the index $q$ labeling the $q$ or the $q'$ direction) we find that the algebra (\ref{p2one})-(\ref{p2three}) contains as a subalgebra the algebra of the $R$-flux model (see \cite{BL}, p. 10) :
 \eqa
& &  [x_a,x_b] = l_s^3 R_{abc} p^c,~~~~[x_a,p^b] = i \hbar \de^{b}_a~ q', ~~~~[p^a, p^b]=0 \\
 & &  [\xtilde_a,\xtilde_b]=0,~~~~~~~~~~~~~ [\xtilde_a,\ptilde^b] = i \hbar \de^{b}_a ~q',~~~~ [\ptilde^a,\ptilde^b]=0
  \ena
\noi where we have rescaled the generators as $x_a \rightarrow l_S x^a$, $p^b \rightarrow {1 \over l_S} p^b$ (same for $\xtilde_a$ and $\ptilde^b$) and $q' \rightarrow i \hbar q'$. The charge $q'$ is central in this subalgebra, and therefore can play the role of the identity. The other nonvanishing commutators involving the extra charge $q$ are
   \eq
   [q,x_a]=x_a,~~~~ [q,p^b]= 2p^b,~~~~  [q,\xtilde_a]=\xtilde_a,~~~~ [q,\ptilde^b]=2\ptilde^b,~~~~ [q,q']=3 q'
   \en
 Note that $q'$ cannot be seen as identity in the whole algebra since it does not commute with $q$. 

Computing the Jacobiator (\ref{p2Jac}) for three generators $x_a,x_b,x_c$ we find it to be proportional to
 $\hbar l^3_s R_{abc}$,  as given in \cite{BL}.  
 
The FDA equations corresponding to the structure constants given in (\ref{sc}) are :
\eqa
 & & d \sigma^a + Q \sigma^a =0 \label{FDA1} \\
 & & d \sitilde^a + Q \sitilde^a=0  \label{FDA2}\\
 & & dQ = 0  \label{FDA3}\\
 & & dB^\bullet + 3 Q B^\bullet + C^\bullet_{abc} \sigma^a \sigma^b \sigma^c =0  \label{FDA4}
 \ena
where $\si^a$ and $\sitilde^a$ are $1$-forms dual respectively to the coordinates $x_a$ and their doubles $\xtilde_a$,
$Q$ is a one-form $U(1)$ gauge field dual to the charge $q$, and $B$ is the $2$-form in a 1-dimensional  $U(1)$ representation
(with a 1-dimensional index $i = \bullet$). It is easy to verify
that $d^2=0$ (the only nontrivial case being $d^2$ on $ B^\bullet$), thus proving that the structure constants (\ref{sc}) indeed satisfy generalized Jacobi identities.

The choice of a 2-form in the FDA's allows to identify the momenta $p^b$ and $\ptilde^b$  as part of the
``new" generators $T^A_\bullet$.

 \sk
 ii)  $B$: $p$-form in the identity representation.
  \eqa
  & & [T_A,T_B] = \c{AB}{C} T_C + {1 \over (p-1)!} \cl{ABC_1...C_{p-1}}{} T^{C_1 ...C_{p-1}}\\
 & & [T_A, T^{B_1...B_{p-1}} ]= -(p-1) \c{AC}{[B_1} T^{CB_2...B_{p-1}]}\\
 & & [T^{A_1 ...A_{p-1}}, T^{B_1...B_{p-1}}] =0
 \ena
 \eq
 J_{ABC}={1 \over (p-2)!} \cl{A_1...A_{p-1} [AB} {}\c{C]D}{A_1} T^{DA_2...A_{p-1}}
 \en
\sk
 iii) $B$: 3-form in a generic $G$ representation:
 \eqa
 & & [T_A,T_B]=\c{AB}{C}T_C + \unmezzo \cl{ABA_1 A_2}{i} T^{A_1A_2}_{i} \\
 & & [T_A,T^{B_1B_2}_i]= \c{Ai}{k} T^{B_1B_2}_k+ 2\c{AC}{[B_1} T^{B_2]C}_i \\
 & & [T^{A_1A_2}_i,T^{B_1B_2}_j] =0
 \ena
 \eq
   J_{ABC}=-( \cl{EA_1[AB} {i}\c{C]A_2}{E}+\unmezzo \cl{A_1A_2 [AB}{j} \c{C]j}{i} ) T_i^{A_1A_2}
 \en
 
The observation in i) on the identification of an extended R-flux model algebra with a dual FDA algebra
  is in fact only a very preliminary step towards understanding and exploiting a possible connection between 
dual FDA algebras and the (double) geometry of flux backgrounds in closed string theory.

\sk
\noi {\bf Acknowledgements}
 \sk
 We thank Ioannis Bakas and Dieter L\"ust for a discussion on 
the nonassociative algebras of ref. \cite{Lust2010,BL}.

\vfill\eject
\end{document}